\documentclass[showkeys, showpacs]{revtex4}
\usepackage{indentfirst}
\begin{document}
\large
\title{\bf Heavy baryonia}
\author{S.M. Gerasyuta}
\email{gerasyuta@SG6488.spb.edu}
\author{E.E. Matskevich}
\email{matskev@pobox.spbu.ru}
\affiliation{Department of Theoretical Physics, St. Petersburg State University, 198904,
St. Petersburg, Russia}
\affiliation{Department of Physics, LTA, 194021, St. Petersburg, Russia}
\begin{abstract}
The relativistic six-quark equations are found in the framework of the
dispersion relation technique. The charmed baryonia $B\bar B$ are constructed
without the mixing of the quarks and antiquarks. The relativistic six-quark
amplitudes of the heavy baryonia are calculated. The poles of these amplitudes
determine the masses of baryonia. 8 masses of charmed baryonia are predicted.
\end{abstract}
\maketitle
In the recent paper \cite{1} the relativistic six-quark equations including
$u$, $d$ quarks and antiquarks are found. The nonstrange baryonia $B \bar B$
are constructed without the mixing of the quarks and antiquarks. Therefore
we did not use the three mesons and the meson plus the tetraquark states.
The contribution of the relativistic amplitudes for these six-quark states
is the other task.

The relativistic six-quark amplitudes of the nonstrange baryonia are calculated.
The poles of these amplitudes determine the masses of baryonia. The dynamical
mixing between the subamplitudes of baryonia is considered. The relativistic
six-quark equations are obtained in the form of the dispersion relations
over the two-body subenergy. The approximate solutions of these equations
using the method based on the extraction of leading singularities of the
amplitude are obtained. We have calculated the mass spectrum of baryonia
and the contributions of subamplitudes to the baryonia amplitude.

The planar diagrams are used; the other diagrams due to the rules of $1/N_c$
expansion are neglected \cite{2, 3, 4}.

The correct equations for the amplitude are obtained by taking into account
all possible subamplitudes. It corresponds to the division of complete system
into subsystems with a smaller number of particles. Then one should represent
a six-particle amplitude as a sum of 15 subamplitudes.

This defines the division of the diagrams into groups according to the
certain pair interaction of particles. We consider the relativistic
generalization of the Faddeev-Yakubovsky approach \cite{5, 6}.

The construction of the approximate solution of six-quark equations is based
on the extraction of the leading singularities of the amplitudes \cite{7}.
The main singularities in $s_{ik}=(m_i+m_k)^2$ are from pair rescattering of
the particles $i$ and $k$. First of all there are threshold square-root
singularities. Also possible are pole singularities which correspond to the
bound states. Apart from two-particle singularities, triangular singularities
and the singularities defining the interactions of four, five and six particles
are considered. Such classification allows us to search the corresponding
solution of equations by taking into account some definite number of leading
singularities and neglecting all the weaker ones.

We consider the approximation which defines two-particle, triangle and
four-,  five- and six-particle singularities.

The contribution of two-particle and triangle singularities are more important,
but we must take into account also the other singularities. Using this
classification, one defines the reduced amplitudes $\alpha_i$ as well
as the Chew-Mandelstam functions in the middle point of physical region of
Dalitz-plot at the point $s_0$.

Such choice of point $s_0$ allows us to replace the integral equations
by the algebraic equations.

We used the functions $I_1$, $I_2$, $I_3$, $I_4$, $I_6$ similar to the paper \cite{8}.

Recently BES Collaboration reported the results on $Y(4630)$ in the
$e^+ e^- \to\Lambda_c \bar \Lambda_c$ process \cite{9}. This hadron is
considered as $\Lambda_c \bar \Lambda_c$ molecular state. The mass and width
of $M=4630\, MeV$ and $\Gamma=92\, MeV$.

The poles of the reduced amplitudes $\alpha_l$ correspond to the bound states
and determine the masses of the charmed baryonia.

The pair quarks amplitudes $qq \to qq$ and $q\bar q \to q\bar q$ are calculated
with the dispersion $N/D$ method with the input four-fermion interaction
\cite{10, 11} with the quantum numbers of the gluon \cite{12, 13}.

In Table \ref{tab1} the calculated masses of charmed baryonia are shown.
The parameters of model are similar to those in the previous paper \cite{14}.

The baryonium $\Lambda_c \bar \Lambda_c$ ($udc\,\, \bar u\bar d\bar c$) is
calculated with the 33 subamplitudes (equations), 24 $\alpha_1$ (for instance,
$\alpha_1^{0^{ud}}$) and 9 $\alpha_2$: $\alpha_2^{0^{ud}0^{\bar u\bar d}}$,
$\alpha_2^{0^{ud}0^{\bar u\bar c}}$, $\alpha_2^{0^{ud}0^{\bar d\bar c}}$,
$\alpha_2^{0^{uc}0^{\bar u\bar d}}$, $\alpha_2^{0^{uc}0^{\bar u\bar c}}$,
$\alpha_2^{0^{uc}0^{\bar d\bar c}}$, $\alpha_2^{0^{dc}0^{\bar u\bar d}}$,
$\alpha_2^{0^{dc}0^{\bar u\bar c}}$, $\alpha_2^{0^{dc}0^{\bar d\bar c}}$.
The isospin is equal to $I=0$ and the spin-parities $J^P=0^-$, $1^-$.
We predict the degeneracy of charmed baryonia (Table \ref{tab1}) with the
mass $M=4410\, MeV$. The similar consideration is used to the states
$\Xi_c \bar \Xi_c$ ($usc\,\, \bar u\bar s\bar c$) with the mass $M=4893\, MeV$.
The model in question the baryonia $\Sigma_c \bar \Sigma_c$ ($uuc\,\, \bar u\bar u\bar c$)
is described with 16 subamplitudes, 12 $\alpha_1$ and 4 $\alpha_2$:
$\alpha_2^{1^{uu}1^{\bar u\bar u}}$, $\alpha_2^{1^{uu}0^{\bar u\bar c}}$,
$\alpha_2^{0^{uc}1^{\bar u\bar u}}$, $\alpha_2^{0^{uc}0^{\bar u\bar c}}$
for the spin-parities $J^P=0^-$, $1^-$. (Baryonium mass $M=4390\, MeV$).
In the case $J^P=2^-$ we considered 15 subamplitudes:
12 $\alpha_1$ and 3 $\alpha_2$: $\alpha_2^{1^{uu}1^{\bar u\bar u}}$,
$\alpha_2^{1^{uu}0^{\bar u\bar c}}$, $\alpha_2^{0^{uc}1^{\bar u\bar u}}$.
(Baryonium mass $M=4422\, MeV$).

The baryonium $\Sigma_c \bar \Sigma_c$ $udc\,\, \bar u\bar u\bar c$ in the case of
spin-parities $J^P=0^-$, $1^-$ is calculated with the 23 subamplitudes (equations),
17 $\alpha_1$ and 6 $\alpha_2$: $\alpha_2^{0^{ud}1^{\bar u\bar u}}$,
$\alpha_2^{0^{uc}1^{\bar u\bar u}}$, $\alpha_2^{0^{dc}1^{\bar u\bar u}}$,
$\alpha_2^{0^{ud}0^{\bar u\bar c}}$, $\alpha_2^{0^{uc}0^{\bar u\bar c}}$,
$\alpha_2^{0^{dc}0^{\bar u\bar c}}$. (Baryonium mass $M=4400\, MeV$).
For the spin-parity $J^P=2^-$ we used the 20 subamplitudes: 17 $\alpha_1$ and
3 $\alpha_2$: $\alpha_2^{0^{ud}1^{\bar u\bar u}}$, $\alpha_2^{0^{uc}1^{\bar u\bar u}}$,
$\alpha_2^{0^{dc}1^{\bar u\bar u}}$. (Baryonium mass $M=4446\, MeV$).

The charmed baryonia $\Omega_c \bar \Omega_c$ $ssc\,\, \bar s\bar s\bar c$ is
described for the spin-parities $J^P=0^-$, $1^-$ using 16 subamplitudes,
12 $\alpha_1$ and 4 $\alpha_2$: $\alpha_2^{1^{ss}1^{\bar s\bar s}}$,
$\alpha_2^{1^{ss}0^{\bar s\bar c}}$, $\alpha_2^{0^{sc}1^{\bar s\bar s}}$,
$\alpha_2^{0^{sc}0^{\bar s\bar c}}$. In the case of spin-parity $J^P=2^-$
15 subamplitudes take into account: 12 $\alpha_1$ and 3 $\alpha_2$:
$\alpha_2^{1^{ss}1^{\bar s\bar s}}$, $\alpha_2^{1^{ss}0^{\bar s\bar c}}$,
$\alpha_2^{0^{sc}1^{\bar s\bar s}}$.

The estimation of theoretical error on the baryonia masses is equal to $1\, MeV$.

\begin{acknowledgments}
The work was carried out with the support of the Russian Ministry of Education
(grant 2.1.1.68.26).
\end{acknowledgments}

\begin{table}
\caption{$qqQ\bar q\bar q\bar Q$, $q=u, d, s$, $Q=c$. Parameters of model: cutoff
$\Lambda=11.0$, $\Lambda_{qc, cc}=8.52$, gluon coupling constant $g=0.314$.
Quark masses $m_q=495\, MeV$, $m_s=770\, MeV$ and $m_c=1655\, MeV$.}
\label{tab1}
\begin{tabular}{|c|c|c|c|c|}
\hline
Quark content & $I$ & $J$ & Baryonium & Mass (MeV) \\
\hline
$uuc\,\, \bar u\bar u\bar c$, $ddc\,\, \bar d\bar d\bar c$; & 0; 2 & 0 &
$\Sigma_c \bar \Sigma_c$, $\Sigma^*_c \bar \Sigma^*_c$ & 4390 \\
$uuc\,\, \bar d\bar d\bar c$, $ddc\,\, \bar u\bar u\bar c$ &  & 1 &
$\Sigma_c \bar \Sigma_c$, $\Sigma_c \bar \Sigma^*_c$, $\Sigma^*_c \bar \Sigma_c$, $\Sigma^*_c \bar \Sigma^*_c$ & 4389 \\
 & & 2 & $\Sigma_c \bar \Sigma^*_c$, $\Sigma^*_c \bar \Sigma_c$, $\Sigma^*_c \bar \Sigma^*_c$ & 4422 \\
\hline
$uuc\,\, \bar u\bar d\bar c$, $ddc\,\, \bar u\bar d\bar c$; & 1 & 0 &
$\Sigma_c \bar \Sigma_c$, $\Sigma^*_c \bar \Sigma^*_c$, $\Sigma_c \bar \Lambda_c$, $\Lambda_c \bar \Sigma_c$ & 4400 \\
$udc\,\, \bar u\bar u\bar c$, $udc\,\, \bar d\bar d\bar c$ & & 1 &
$\Sigma_c \bar \Sigma_c$, $\Sigma_c \bar \Sigma^*_c$, $\Sigma_c \bar \Lambda_c$,
$\Sigma^*_c \bar \Sigma_c$, $\Sigma^*_c \bar \Sigma^*_c$, $\Sigma^*_c \bar \Lambda_c$,
$\Lambda_c \bar \Sigma_c$, $\Lambda_c \bar \Sigma^*_c$ & 4400 \\
 & & 2 & $\Sigma_c \bar \Sigma^*_c$, $\Sigma^*_c \bar \Sigma_c$, $\Sigma^*_c \bar \Sigma^*_c$,
$\Sigma^*_c \bar \Lambda_c$, $\Lambda_c \bar \Sigma^*_c$ & 4446 \\
\hline
$udc\,\, \bar u\bar d\bar c$ & 0 & 0 & $\Sigma_c \bar \Sigma_c$, $\Sigma_c \bar \Lambda_c$,
$\Lambda_c \bar \Lambda_c$, $\Sigma^*_c \bar \Sigma^*_c$, $\Lambda_c \bar \Sigma_c$ & 4410 \\
 & & 1 & $\Sigma_c \bar \Sigma_c$, $\Sigma_c \bar \Lambda_c$, $\Sigma_c \bar \Sigma^*_c$,
 $\Sigma^*_c \bar \Sigma_c$, $\Sigma^*_c \bar \Lambda_c$, $\Sigma^*_c \bar \Sigma^*_c$,
$\Lambda_c \bar \Sigma_c$, $\Lambda_c \bar \Lambda_c$, $\Lambda_c \bar \Sigma^*_c$ & 4410 \\
\hline
$usc\,\, \bar u\bar s\bar c$, $dsc\,\, \bar d\bar s\bar c$; & 0; 1 & 0 &
$\Xi_c \bar \Xi_c$, $\Xi^*_c \bar \Xi^*_c$ & 4893 \\
$usc\,\, \bar d\bar s\bar c$, $dsc\,\, \bar u\bar s\bar c$; & & 1 &
$\Xi_c \bar \Xi_c$, $\Xi_c \bar \Xi^*_c$, $\Xi^*_c \bar \Xi_c$, $\Xi^*_c \bar \Xi^*_c$ & 4893 \\
\hline
$ssc\,\, \bar s\bar s\bar c$ & 0 & 0 & $\Omega_c \bar \Omega_c$, $\Omega^*_c \bar \Omega^*_c$ & 5254 \\
 & & 1 & $\Omega_c \bar \Omega_c$, $\Omega_c \bar \Omega^*_c$, $\Omega^*_c \bar \Omega_c$,
 $\Omega^*_c \bar \Omega^*_c$ & 5252 \\
 & & 2 & $\Omega_c \bar \Omega^*_c$, $\Omega^*_c \bar \Omega_c$, $\Omega^*_c \bar \Omega^*_c$ & 5278 \\
\hline
\end{tabular}
\end{table}

\end{document}